\documentclass[twocolumn,secnumarabic,amssymb, nobibnotes, aps, pra,,showpacs,preprintnumbers,amsmath,amssymb,groupaddress]{revtex4-1}
\usepackage{amsmath}
\usepackage{amssymb}
\usepackage{graphicx}
\usepackage{epstopdf}
\usepackage{subfigure}
\usepackage{epstopdf}
\usepackage{braket}
\usepackage{bbold}
\usepackage{amsmath}
\usepackage{bm}
\usepackage{subfigure}
\usepackage{tabularx}
\usepackage{multirow}
\begin{document}

\title{Optimal control of fast and high-fidelity quantum gates with
  electron and nuclear spins of a nitrogen-vacancy center in diamond}

\author{Yi Chou}
\author{Shang-Yu Huang}
\author{Hsi-Sheng Goan}
\email{goan@phys.ntu.edu.tw}
\affiliation{Department of Physics and Center for Theoretical Sciences,
National Taiwan University, Taipei 10617, Taiwan}
\affiliation{Center for Quantum Science and Engineering,
National Taiwan University, Taipei 10617, Taiwan}

\date{\today}
\begin{abstract}
A negatively charged nitrogen vacancy (NV) center
in diamond has been recognized as a good solid-state qubit. 
A system consisting of the electronic spin of the NV center and
hyperfine-coupled nitrogen and additionally nearby 
carbon nuclear spins can form 
a quantum register
of several qubits
 for quantum information processing or 
as a node in a quantum repeater. 
Several impressive experiments on the hybrid electron
and nuclear spin register have been reported, but 
fidelities achieved so far are not yet at oor below the thresholds required for
fault-tolerant quantum computation (FTQC).
Using quantum optimal control theory based on the Krotov method,
we show here that fast and high-fidelity single-qubit and two-qubit
gates in the universal quantum gate set for
FTQC, taking into account the effects of the leakage state, nearby 
noise qubits and distant bath spins, 
can be achieved with errors less than those
required by the threshold theorem of FTQC. 
\end{abstract}

\pacs{03.67.Lx,03.65.Yz,76.30.Mi,02.30.Yy}

\maketitle

\section{INTRODUCTION}
Nitrogen vacancy (NV) centers in diamond have many remarkable
properties. For example, the spins of NV centers have relatively long relaxation
and coherence
time (even at room temperature)
\cite{Balasubramanian2009,Jarmola2012,Ishikawa2012,Fang2013,Bar-Gill2013,Maurer2012},
and the electron spin triplet ground state can
be initialized, manipulated and read out with microwaves and lasers. 
\cite{Jelezko2002,Jelezko2004,Epstein2005}. 
These exceptional properties
make the NV center(s) a promising system for sensitive magnetic field
and weak signal
sensing \cite{Balasubramanian2008,Maze2008,Acosta2009,Kolkowitz2012,Taminiau2012,Zhao2012,Staudacher2013,Mamin2013,Maletinsky2012,Loretz2013,Cai2014}, 
bio-marking tracking \cite{Chung06,Fu2007,Chao2007,Neugart2007}
and
quantum information processing
\cite{Beveratos2002,Beveratos2002L,Jelezko2004,Jelezko2004CNOT,Childress2005,Childress2006,Jiang2007,Barrett2005,Dutt2007,Neumann2008,Fuchs2009,Cappellaro2009,Neumann2010,Shi2010,Fuchs2011,Xu2012,van-der-Sar2012,Liu2013,Shim2013,Dolde2013,Zhang2014,Rong2014,Taminiau2014,Waldherr2014,Dolde2014,Scheuer2014,Scharfenberger2014}.

Quantum gate operations in a quantum register of individual electron
spin or/and nearby 
individual nuclear spins associated with a NV center in diamond 
 have been demonstrated experimentally
\cite{Jelezko2004CNOT,Dutt2007,Neumann2008,Fuchs2009,Cappellaro2009,Neumann2010,Shi2010,Fuchs2011,Xu2012,van-der-Sar2012,Liu2013,Shim2013,Dolde2013,Taminiau2014}. However,
the gate fidelities 
in theses experiments or studies are limited to certain values
because the pulses sequences
to perform the gates even in the ideal unitary case 
are not optimally designed and come with some sorts of approximation. 

There have been schemes proposed for protecting quantum
gates of NV center spins 
from decoherence, based on dynamical decoupling (DD) protocols
or/and dynamically corrected gate (DCG)
\cite{deLange01102010,van-der-Sar2012,Wang2012,Zhao2012PRB,Zhang2014,Rong2014}.  
Experimental realizations of noise-resilient or decoherence protected
 quantum gates on NV centers have been reported \cite{Liu2013}. 
So far, only the single-qubit gates are shown to be of high-fidelity. 
For example, the fidelity of a dynamical-decoupling-protected $X$ gate
is shown to be about
0.985 for a gate duration of $35.5$ $\mu$s
\cite{Zhang2014} and the fidelity of a \textsc{SUPCODE}
$\pi/2$ gate is about 0.9961 with a gate time of  $5.063$ $\mu$s
\cite{Rong2014}.   
However, these gate times are much longer than those of unprotected
gates, and how to practically implement the different protected
gates for different qubits in parallel
in a many-qubit register is not clear.

An alternative approach to realize high-fidelity quantum gate is
through the quantum optimal control (QOC)
\cite{Rabitz88,Tannor92,Kosloff02,Tannor2002,Xu2004,Khaneja05,Sporl07,Tsai2009,Tannor11,Montangero2007,Nielsen08,potz2008,potz2009,Jirari2009,Rebentrost2009,Hwang2012,Tai2014,Huang2014}. 
A recent study \cite{Scharfenberger2014}
investigated the theoretically achievable fidelities when coherently controlling
an effective three-qubit system consisting of a negatively charged ($^{15}$NV$^{-}$) center in diamond with an additional nearby
carbon $^{13}$C nuclear spin. The results in this study indicates that by using square and two frequency component radio and microwave frequency
pulses, the best single-qubit gate fidelity 
is less than 98$\%$ and the multi-qubit gate
fidelities are somewhat lower than that. 
It was thus suggested that to
reach the fidelity threshold(s)
predicted by current models of fault-tolerant quantum computation
(FTQC) \cite{Aliferis2009,Wang11,Fowler12L,Fowler12}, going beyond the square-pulse
paradigm and using pulse-shaping techniques like optimal control is
required \cite{Scharfenberger2014}. 
The QOC theory has
been applied to NV center based quantum information
processing \cite{Waldherr2014,Dolde2014,Scheuer2014}.  Reference
\onlinecite{Scheuer2014} considered only the electron spin and designed
fast single-qubit gates using the chopped random basis quantum optimization
algorithm without resorting to the standard  rotating-wave
approximation condition. 
References \onlinecite{Waldherr2014} considered a system of
a NV center's electron spin and nitrogen
$^{14}$N nuclear spins as well as coupled carbon $^{13}$C nuclear
spins, forming a small quantum register, and 
used the gradient ascent pulse engineering (GRAPE) optimization
algorithm \cite{Khaneja05,Sporl07,Tsai2009} to  
perform  phase-flip quantum error correction on three qubits of one
$^{14}$N nuclear spin and two $^{13}$C nuclear spins. 
Reference
\onlinecite{Dolde2014} performed 
 quantum gates and generated entangled states for two proximal NV
 centers in diamond using the GRAPE
 optimization algorithm. 
With the help of QOC, unwanted off-resonance transitions
or crosstalk, and unwanted dipolar couplings between the spins of the two proximal NV
 centers were significantly suppressed.  
However, in these QOC studies and experiments \cite{Waldherr2014,Dolde2014,Scheuer2014},
the maximum hyperfine interaction strength between the
NV electron spin and either the nitrogen nuclear spin or carbon
nuclear spin is only about a few MHz. 
This is different from the case studied in Ref.~\onlinecite{Scharfenberger2014} 
where the hyperfine interaction of the $^{15}$NV$^{-}$ is about 3MHz,
while the hyperfine interaction with the nearest neighbor 
carbon $^{13}$C  can be larger than 100MHz. 
A large hyperfine interaction potentially leads to fast quantum
gate operations in the hybrid spin register.
It is, however, this large hyperfine interaction that limits the
maximum fidelity that can be achieved in the square-pulse paradigm \cite{Scharfenberger2014}. 
Furthermore, these studies  \cite{Waldherr2014,Dolde2014,Scheuer2014}
do not take the
decoherence effect from the surrounding distant bath spins into the
optimization consideration
when constructing their QOC gates.

In this paper, we present a detailed QOC study
based on the Krotov optimization method \cite{krotov1996,Tannor92,Kosloff02,Tannor2002,Montangero2007,Nielsen08,Tannor11,Hwang2012,Tai2014,Huang2014}
for
 single-qubit and two-qubit gates of a hybrid electron
 and nuclear spin register of a NV center in diamond taking into
account the effects of leakage state, nearby noise spins and a bath  
of distant nitrogen spins or/and $^{13}$C nuclear spins randomly
distributed in the diamond lattice.
In our model, a nuclear spin of $\mathrm{^{13}C}$ which is 
in the first coordination
shell (nearest neighbor) around the $\mathrm{^{15}NV^-}$ center
and has strong hyperfine interaction with
the NV center electron spin is considered (see Fig.~\ref{fig:model} for
a schematic illustration).
The Krotov optimization method we employ has several appealing advantages
over the gradient methods \cite{krotov1996,Tannor92,Kosloff02,Tannor2002,Nielsen08,Tannor11}: (a) monotonic increase
of the objective with iteration number, (b) no requirement for
a line search, and (c) macrosteps at each iteration. 
Quantum gates constructed via our QOC scheme with 
experimentally available or realistic parameters  
are all with 
very fast speed and very high fidelity. 
Setting the gate operation times for our single-qubit $X$ gate and
$Z$ gate performed on
the electron spin to be $10$ ns, we obtain corresponding gate infidelities or
errors to be $3.9\times 10^{-5}$ and $6.0\times 10^{-4}$,
respectively. The two-qubit controlled-Not ({\sc CNOT}) gate performed on the NV center
electron spin and a proximal 
$^{13}$C nuclear spin can be operated within $50$ ns with an
infidelity or error of about $4.3\times 10^{-4}$ even in the presence of a
host $^{15}$N   
noise nuclear spin and an additional spin bath (environment) with a
wide range of decoherence parameters.  

This paper is organized as
follows. We briefly describe the model Hamiltonian of the
NV-center-based hybrid spin register we consider in Sec.~\ref{sec:model}.  
To incorporate the effect of distant nuclear spins which form a spin
bath or environment on the dynamics of the system qubits, we use the
open-system master-equation approach and the description of this
approach is presented in Sec.~\ref{sec:ME}. 
In Sec.~\ref{sec:QOCT}, we define the infidelity or error function 
to measure how well the gate operations of our system qubits deviate the
ideal target gates in the presence of nearby noise qubits (spins) and
a spin bath.
The QOC algorithm based on the Krotov method 
is also briefly described here. 
In Sec.~\ref{sec:results}, we explore the application of the 
QOC for the implementations
of $Z$-gate, $X$-gate and also the gates in the universal discrete 
quantum gate set for 
FTQC for the
hybrid spin register.
Comparison to the
traditional approach of implementing quantum gates and the effect of
spin bath on the QOC gate
operations are also presented.
Finally, a conclusion is given in Sec.~\ref{sec:conclusion}.

\section{Model Hamiltonian}\label{sec:model}

\begin{figure}
\mbox{
\subfigure[]{\includegraphics[scale=0.15]{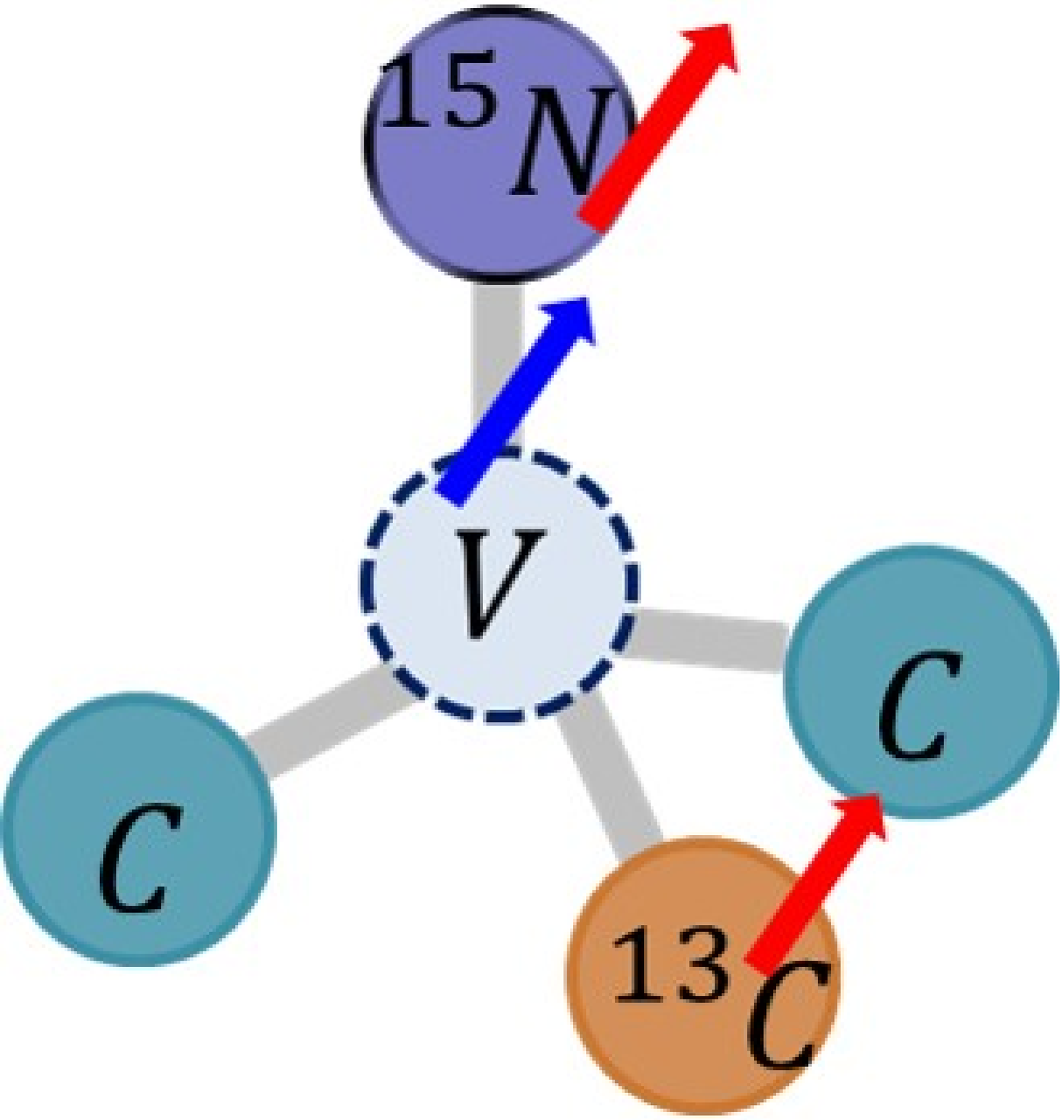}}\hspace{1cm}
\subfigure[]{\includegraphics[scale=0.15]{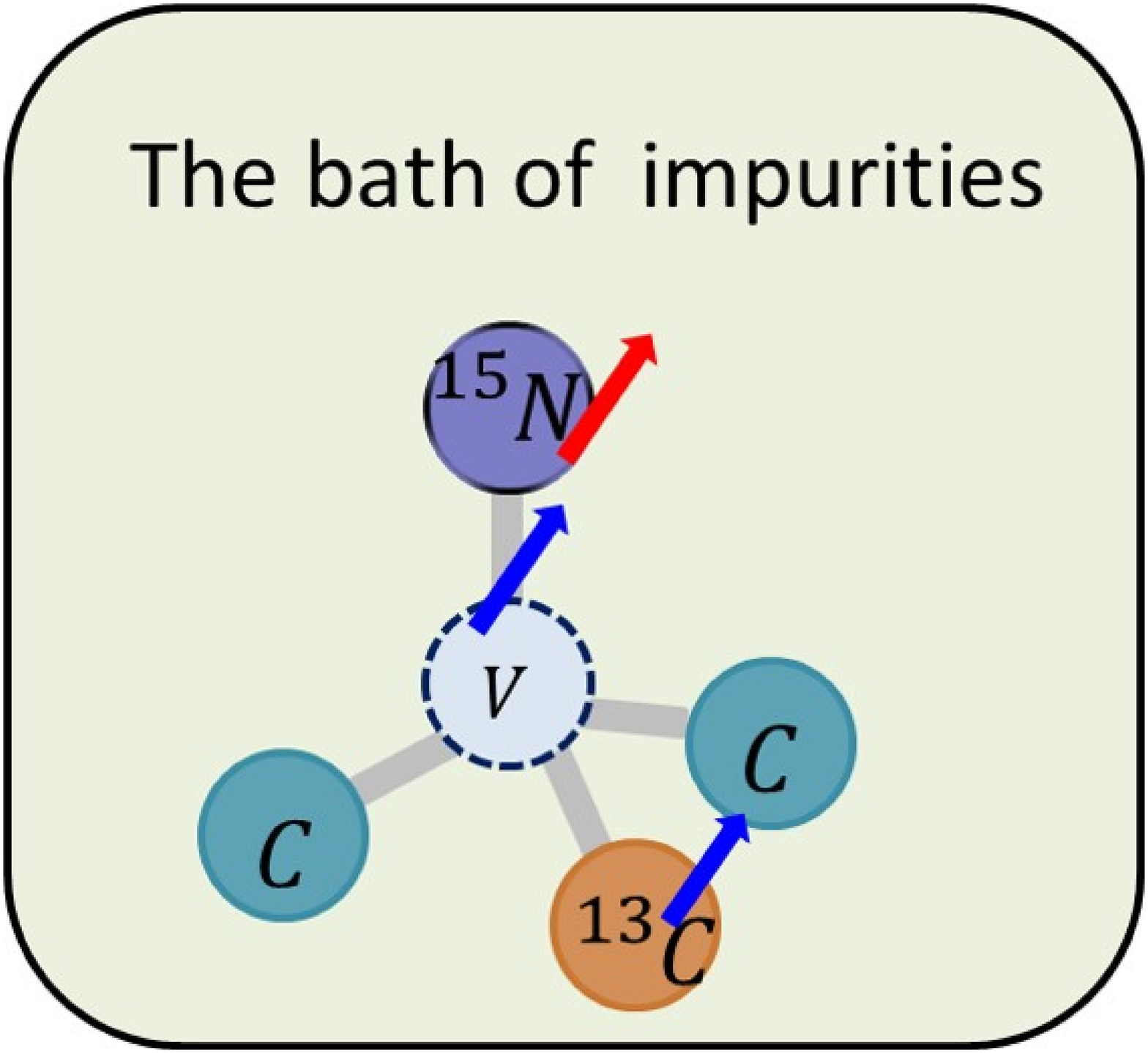}}}
\caption{\label{fig:model} (Color online) 
Schematic illustration of a negatively charged  $\mathrm{^{15}NV^-}$
center (V denoting a vacancy site) with one
  nearest neighbor $\mathrm{^{13}C}$ atom. (a) For single-qubit
  operation, the electron spin of the NV center (in blue) is the
  system qubit while the $\mathrm{^{15}N}$ and $\mathrm{^{13}C}$
  nuclear spins are regarded as noise qubits (in red). (b) For two
  qubit operations, the electron spin and the $\mathrm{^{13}C}$
  nuclear spin are regarded as the system qubits (in blue), while
  the  $\mathrm{^{15}N}$  
  nuclear spin is treated as a noise qubit (in red). The shaded area
  denotes additional
  distant bath spins.}
\end{figure}

We consider a negatively charged NV center associated with a
$\mathrm{^{15}N}$ nucleus (i.e., $\mathrm{^{15}NV^-}$) in diamond . The
electronic structure of the NV center has a spin-triplet ground state
$S = 1$
with a zero-field splitting
$\bigtriangleup = \mathrm{2.87\cdot 2\pi\, GHz}$ between the $m_s=0$ and
$m_s=\pm 1$ levels. Note that the quantization axis of this splitting
is along the symmetry axis of the NV center, which we take as the $z$
axis. The $\mathrm{^{15}N}$ carries a nuclear spin
$I=\frac{1}{2}$. Also, we consider a  $\mathrm{^{13}C}$ atom occupies one of
the nearest position around the NV center, and other nuclear spins further
away are regarded as a spin bath. A schematic structure of our system is shown
in Fig.\ref{fig:model}. Applying a static magnetic field $B$
along the $z$ axis splits the levels $m_s=-1 $ and $m_s=+1$. 
In our study, we consider all the three electron spin levels, i.e.,
the $m_s=0$ and $m_s=\pm 1$ levels,
 choose
the spin levels $m_s=0$ and $m_s=-1 $ to be the two computational states of our
electron spin qubit
and treat the $m_s=+1$ state as an ancilla or a leakage
state. 
The NV center electron spin
is coupled to
the proximal $\mathrm{^{15}N}$ and $\mathrm{^{13}C}$
nuclear spins and an additional spin bath.

It has been shown that the coupling of a spin bath of 
 distant nuclear spins to a NV center electron spin can be modeled
 through classical magnetic field 
 noise that causes decoherence by imprinting a random phase on the NV center
electron spin \cite{Reinhard2012}.
This semiclassical noise model emerges as the weak-coupling
limit of quantum-mechanical entanglement-induced decoherence.
The condition for this to be valid is when the
electron-nuclear spin couplings quantified
by the magnetic fields $B_{{\rm NV}.n}$ of the NV center electron spin
at the sites of the nuclear 
spins are much smaller than the externally applied magnetic
field $B_z$ (i.e.,$B_{{\rm NV}.n} \ll B_z$) \cite{Reinhard2012}. 
In our study, we apply a relatively strong background magnetic field
$B_z$ to the spin register. 
We thus treat the coupling of the spin bath to the NV center electron spin
to be approximated as a classical
 random field $B(t)$ which may depend on time
acting on the $z$-component of the NV center electron spin, a pure
dephasing model
\cite{deLange01102010,PhysRevB.87.115122,1367-2630-14-11-113023,PhysRevLett.107.150503}.
The couplings of the nearby or neighboring nuclear spins to the
NV center electron spin 
are, however, treated quantum-mechanically.

The total Hamiltonian of the system we consider can be written as follows.   
\begin{align}
&H=H_{0}+H_{cx}+H_{cy}+H_{eB}
\label{H_total}\\
&H_{0}=\bigtriangleup S_z^2-\gamma_e B_z S_z-\gamma_C B_z I_{Cz}- \gamma_N B_z I_{Nz} \nonumber \\
&\hspace{1cm} +A_{\parallel}^{eC} S_z I_{Cz} + A_{\perp}^{eC} \left( S_x I_{Cx} +S_y I_{Cy}\right)\nonumber\\
&\hspace{1cm} +A_{\parallel }^{eN} S_z I_{Nz} +A_{\perp}^{eN} \left( S_x I_{Nx} +S_y I_{Ny}\right), \\
&H_{cx}(t)=B_x(t) \left( -\gamma_e S_x -\gamma_C I_{Cx} -\gamma_N I_{Nx} \right),\\
&H_{cy}(t)=B_y(t) \left( -\gamma_e S_y -\gamma_C I_{Cy} -\gamma_N I_{Ny} \right),\\
&H_{eB}=S_z\tilde{B}(t),
\end{align}
where $S_{i}$ is the spin-1 operators for the
NV center electron spin and
$I_{Ci}$ and $I_{Ni}$ are the spin-1/2 operators for the $\mathrm{^{13}C}$ and $\mathrm{^{15}N}$
nuclear spins, respectively. $\gamma_e= \mathrm{-2.8\cdot 2\pi \, MHz\cdot
  G^{-1}}$, $\gamma_C=\mathrm{0.00107\cdot 2\pi \, MHz\cdot G^{-1}}$ and
$\gamma_N=\mathrm{-0.43\cdot 2\pi \, kHz\cdot G^{-1}}$ are the
gyromagnetic ratios for the electron, $\mathrm{^{13}C}$ and
$\mathrm{^{15}N}$ spins, respectively \cite{Zhao2012PRB}.
$A_{\parallel}^{eC}=A_{\perp}^{eC}= \mathrm{127\cdot 2\pi MHz}$ is the
hyperfine coupling between the electron and the $\mathrm{^{13}C}$ nuclear
spin \cite{Shim2013}, and $A_{\parallel}^{eN}=A_{\perp}^{eN}= \mathrm{3.03\cdot 2\pi
  MHz}$ is that between the electron and the $\mathrm{^{15}N}$ nuclear
spin \cite{Rabeau2006,Wang2012}. The magnetic fields $B_x(t)$ and  $B_y(t)$ are time-dependent external control fields.
The random field $\tilde{B}(t)$ represents the effect of the bath
spins with correlation function  \cite{deLange01102010,PhysRevB.87.115122,1367-2630-14-11-113023,PhysRevLett.107.150503}
\begin{eqnarray}
C(t,t')&=&\langle \tilde{B}(t)\tilde{B}(t')\rangle \nonumber \\
&=& b^2 e^{-\mid t-t'\mid /\tau_c},
\label{eq:bathCF}   
\end{eqnarray}
where $b$ that can be regarded as the field inhomogeneity in the
effective semi-classical random
noise model
is associated with the average coupling strength between the
electron spin and the spin bath,
and $\tau_c$ is the correlation time of
the spin bath. 

We will consider first the case where the quantum gate operations of
the system qubits are influenced by the interactions with 
nearby noise qubits with a low number of degrees of freedom \cite{Scharfenberger2014,0953-4075-40-9-S06,doi:10.1080/09500340701639615}. 
In this case, the most general approach is to describe the dynamics of
both the system and noise qubits and their interactions in Hamiltonian
unitary approach and then perform the QOC calculations for the quantum
gate operations. As the number of degrees of freedom of the noise qubits
increases, the computation of this approach becomes
expensive and challenging. 
Our strategy is that we treat the spins which are close to and 
have large hyperfine
interactions with the system qubit(s) as the noise qubits and include
them and their interactions into the unitary Hamiltonian, and then 
take the many randomly distributed 
distant spins as a  
spin bath (environment) whose average effect on the dynamics of the
system qubit(s) is obtained
by tracing out the environment degrees of freedom (or more precisely by
performing an ensemble average over the classical random noise)
using the master equation approach.
 of the reduced density matrix. 
So our QOC treatment can simultaneously deal with the
effects of leakage states, a few nearby noise qubits and a (spin) bath.

\section{Quantum master equation}\label{sec:ME}
Since the ensemble average effect of the
semi-classical random noise treatment of the spin bath
 in the absence
of the external control fields can be modeled by a 
pure dephasing open system model \cite{PhysRevA.82.012111}, 
we employ the perturbative time-local non-Markovian master equation
to describe the coherent control and decoherence dynamics of the NV
center electron qubit in our QOC study. 
Thus, following the the
standard perturbation theory with the Born approximation, we can write
the time-local non-Markovian master equation for the reduced system
density matrix 
as \cite{raey,welack:044712,kleinekathofer:2505} 
\begin{equation}
 \dfrac{d \rho(t)}{d t}=\mathcal{L}_s \rho(t)+\left[ \mathcal{L}_z D(t) \rho(t)+h.c.\right],  \label{eq:TL}
\end{equation}
where $\mathcal{L}_s=\dfrac{-i}{\hbar}\left[ H_S(t),\bullet\right]$ and $\mathcal{L}_z=\dfrac{-i}{\hbar}\left[ S_z,\bullet\right]$, and the dissipator $D$ can be written as \cite{raey,kleinekathofer:2505}
\begin{equation}
D(t)=\dfrac{-i}{\hbar}\int_0^t C(t-t') \mathcal{U}_s(t,t')S_z dt',
\label{eq:D_TL} 
\end{equation}
where the propagator superoperator
$\mathcal{U}_s(t,t')=T_{+}e^{\int_{t'}^t 
  \mathcal{L}_s(\tau) d\tau}$ with $T_{+}$ denoting the time-ordering
operator.
The symbol $h.c.$ in Eq.~(\ref{eq:TL}) denotes the hermitian conjugate
of its previous term. 
To solve the master equation (\ref{eq:TL}) directly, one would
need to evaluate the term $\mathcal{U}_s(t,t')S_z$ and then perform
the integration for the dissipator $D(t)$ of Eq.~(\ref{eq:D_TL}). 
This procedure is often numerically inefficient.
Instead, since the bath correlation function 
given in Eq.~(\ref{eq:bathCF}) is in an exponential form, 
one can take the time derivative on Eq.~(\ref{eq:D_TL}) and 
obtain straightly the differential equation for the dissipator
$D(t)$ as
\begin{equation}
\dfrac{d}{dt} D(t)=-i b^2 S_z
+\left(\mathcal{L}_s(t)-\frac{1}{\tau_c}\right)D(t). 
\label{eq:dD_TL}
\end{equation} 
Equations (\ref{eq:TL}) and (\ref{eq:dD_TL}) form a coupled set of inhomogeneous
differential equations, and one can use the Runge-Kutta method to
solve these  equations numerically. 

For the convenience of numerical computation, 
we transform the density matrix $\rho$ into a column vector
$\vec{\rho}$ and in this case  Eq.~(\ref{eq:TL}) becomes
 $ \dot{\vec{\rho}}(t)= \Lambda(t)\vec{\rho}(t)$, where $\Lambda(t)$
 is the corresponding operator associated with Eq.~(\ref{eq:TL}) in
 column vector representation. 
It can be shown that the effective propagator $ U (t)$ defined by the relation
$\vec{\rho}(t)=U(t)\vec{\rho(0)}$ satisfies:
\begin{equation}
\dot{U}(t)=\Lambda(t)U(t) \quad \mbox{with~} U(0)=I_{\mathcal{N}}, 
\label{eq:motion_open_U}
\end{equation}
where  $\mathcal{N}$ is the dimension of $U$ and
$I_{\mathcal{N}}$ denotes the $ \mathcal{N} \times \mathcal{N}$
identity matrix in the operator dimension of $U$.

\section{GATE ERROR AND OPTIMAL CONTROL ALGORITHM}\label{sec:QOCT}

We describe below briefly the error function of the gates and the
optimal control algorithm based on the Krotov method that we adopt for our
calculations. 
The noise qubits are regarded as an effective small environment
interacting with the
system qubits that serve as a register for quantum
information processing.
The implementation of quantum gates in the presence of
only a few
noise qubits using the Hamiltonian unitary approach for QOC
calculations has been
investigated \cite{0953-4075-40-9-S06,doi:10.1080/09500340701639615}. 
Here besides a few noise qubits, the leakage state and the spin
bath are also considered. 
We thus define the
error function $K$ as the distance measure between the associated
propagator $PU(T)$ at the final time $T$ of the composite system and the
target gate $G$ in the column vector representation as follows \cite{0953-4075-40-9-S06,doi:10.1080/09500340701639615}: 
  \begin{equation}
  K=\lambda_N \min_{\Phi}\left\{ \parallel
    PU-G\otimes\Phi \parallel^2_F \mid \Phi^\dagger \Phi=I_{n_B}
  \right\}.
 \label{eq: distance_measure}
  \end{equation}
Here $P$
 denotes a projection operator to project the propagator $U$ 
onto the composite subspace 
 spanned by the tensor product of the system qubit computational basis
states and the noise qubit basis
states \cite{potz2009,arxiv0606064},  $n_B$  is
the dimension of the Hilbert space for the noise qubit subsystem,
and $\Phi$ is an arbitrary unitary
acting only on the noise qubit Hilbert space.
The symbol $\parallel
 \cdot \parallel_F$ stands for the Frobenius matrix norm: 
$\parallel A \parallel_F=(\mathbf{Tr}A^\dagger A)^{1/2}$, 
and $\lambda_N =\frac{1}{2N}$ 
is a normalization factor to
 keep the value of $K$ in the range [0,1] with $N$ the dimension of $PU$. 
The squared norm of $\parallel PU-G\otimes\Phi\parallel_F^2$ is
 minimized over the set of all possible unitary $\Phi$ because we do not care
 what the evolution of the noise qubits is 
as long as the target gate $G$ to be implemented can be achieved.  
The fidelity defined as 
$F=1-K$ with $K$ given in Eq.~(\ref{eq: distance_measure})
is introduced to 
measure how well $PU(T)$ approaches the target gate $G$ at the final
gate time $T$. 
The error function similar to $K$ of Eq.~(\ref{eq: distance_measure})
for a closed composite system consisting of
the system qubits and a few noise qubits 
was defined
and simplified to a computable form in Ref.~\onlinecite{arxiv0606064}. 
Here, we generalize the expression of the error
 function $K$ for an open system with leakage states and/or a
spin bath in a computable form as (cf.  \cite{arxiv0606064}) 
\begin{equation}
 K=\frac{1}{2}+\frac{1}{2N}\mathbf{Tr}[(PU)^\dagger
PU]-\frac{1}{N}\mathrm{Re}\mathbf{Tr}\sqrt{Q^\dagger Q}.  
\label{eq:error_leakage} 
\end{equation}
Here in Eq.~(\ref{eq:error_leakage})
\begin{equation}
  \label{eq:Q}
Q=\sum_{i,j=1}^{n_S} G_{ij}^{\ast} (PU)_{\left( ij\right)},  
\end{equation}
where $PU_{\left( ij\right) }$ are $n_B\times n_B $ matrix partitions of
$PU$ in the computational state basis of the system qubits, and
$ G_{ij}$ are the scalar matrix elements of the target operation $G$
\cite{arxiv0606064}.  
Note that because the projected propagator $PU$ is no
longer unitary when the effect of the leakage states or/and the 
open system environment is considered,
the second term in the error function (\ref{eq:error_leakage})
takes the form of $\mathbf{Tr}[(PU)^\dagger
PU]/({2N})$.  This term approaches $(1/2)$ for an
ideal, closed composite system when $PU\to U$ is unitary,
and in this case the error function $K$ (\ref{eq:error_leakage})
reduces to that of Ref.~\onlinecite{arxiv0606064}.

In realistic control problems, it is desirable that the
optimal control sequence can provide the highest quality
(fidelity) with minimum energy consumption. Therefore, we
define the objective function to be maximized for our optimal control problem
as
\begin{equation}
J=F-\int_0^T \dfrac{\lambda(t)}{2}\left[
  \varepsilon(t)-\tilde{\varepsilon}(t)\right]^2 dt,
\label{eq:J}
\end{equation}
with a weighting function $\lambda(t)>0$ adjusted and chosen
empirically.
Here the reference field 
$\tilde{\varepsilon}(t)$ is chosen to be the control field
of the previous iteration \cite{Tannor92,Kosloff02}. 
In this case, when the iterative procedure approaches the
optimal solution, the change in the control field is minimal or
vanishing. Therefore, this choice of the reference field 
$\tilde{\varepsilon}(t)$ 
ensures that the iterative method is found to increase the total
objective $J$ of Eq.~(\ref{eq:J}) by increasing the gate fidelity $F=1-K$
rather than reducing the total control pulse energy.

The iterative algorithm of the Krotov method used in our optimal
control study of implementing quantum gates is
described briefly as follows
\cite{krotov1996,Tannor92,Kosloff02,Tannor2002,Nielsen08,Tannor11,Hwang2012,Tai2014}. (1)
An initial guess for the values of the control parameters
$\varepsilon_i^0(t)$ is 
randomly chosen [here the control parameters $\varepsilon_i(t)$ can
be the externally applied ac magnetic fields $B_i(t)$ with
$i=x,y$ components]. (2) The evolution propagator $ U[\varepsilon_i^0(t)] $ is
evolved forward in time until $t=T$ using Eq.~(\ref{eq:motion_open_U}). (3) An
auxiliary function $\mathcal{B}[\varepsilon_i^j (t)]$, $j = 0$ for the first
iteration, is evolved backward in time until $t = 0$ using the equation
of motion $\dot{\mathcal{B}}(t)= -\mathcal{B}(t)\Lambda(t)$ and
the boundary condition $\mathcal{B}(T)=-\dfrac{dK}{d(PU)}$. The
explicit form of $\dfrac{dK}{d(PU)}$ can be found in the Appendix
\ref{sec:Append}. (4) The 
updated propagator $U[\varepsilon_i^{j+1}(t)]$ is propagated again forward in
time, while the control parameter $\varepsilon_i(t)$ is updated iteratively with
the rule $\varepsilon_i^{j+1}(t)=\varepsilon_i^j (t)+ \dfrac{1}{\lambda(t)}
\mathrm{Re}\left\{ \mathbf{Tr}[\mathcal{B}^j (t)\dfrac{\partial
    \Lambda(t)}{\partial \varepsilon_i(t)} U^{j+1}(t)]\right\}.$ (5) Steps (3)
and (4) are repeated until either the error $K^j$ is smaller than a
preset value or the ratio of the improved error at the next iteration,  
$\frac{K^{j}-K^{j+1}}{K^j}$, is rather small. After a sufficient number
of iterations, the algorithm converges and the fidelity $F$ in the 
objective function of Eq.~(\ref{eq:J}) reaches asymptotically a
maximum value of $F_{\rm max}$.

\section{RESULTS AND DISCUSSION}\label{sec:results}

\subsection{Optimal control for single-qubit gates}

The target quantum single-qubit gates considered in our investigation are
$Z$-gate, $X$-gate, Hadamard gate ($H$ gate), phase gate and $\pi/8$ gate on the
NV center electron spin. 
We restrict our investigation to have a control magnetic field
only in the $x$-direction, i.e., $B_x(t)$, for the implementation of
these single-qubit 
gates. A static magnetic field $B_z = \mathrm{500 \; G}$ is applied to
split the  $m_s = \pm 1$ states of an NV center electron spin; 
the $m_s = 0$ state and $m_s = -1$ state are chosen as the system
qubit states, and the $m_s=1$ state is treated as a leakage state.
We will investigate the case where there are two nearby noise qubits, 
the $\mathrm{^{13}C}$ nuclear spin and $\mathrm{^{15}N}$ nuclear spin,
and the strength of the hyperfine interaction between the NV center
electron spin and the nearest-neighbor $\mathrm{^{13}C}$ nuclear spin
is comparable to the 
Zeeman splitting of system qubit states.  

A quantum $Z$-gate in the absence of the noise qubits can be realized
by free evolution of the system qubit with high fidelity (or error
smaller than $10^{-8}$ ).  
However, the Z-gate errors $K$ obtained by free evolution
taking the energy shift by the $\mathrm{^{13}C}$ and $\mathrm{^{15}N}$
noise qubites into account at operation times of $\mathrm{0.34\;ns}$,
$\mathrm{7.45\;ns}$ and $\mathrm{15.2\;ns}$
are all greater than $6.0 \times 10^{-3}$.  
On the other hand, using optimal control
method with an extra control field $B_x(t)$ gives, for example, a
$Z$-gate error $K$ 
for an operation
time of $\mathrm{10\;ns}$ (or $0.01\mu$s) 
to be $6.0 \times 10^{-4}$, at least one order of magnitude better than that by
the free evolution.

\begin{figure}
\centering
\subfigure[]{\includegraphics[scale=0.5]{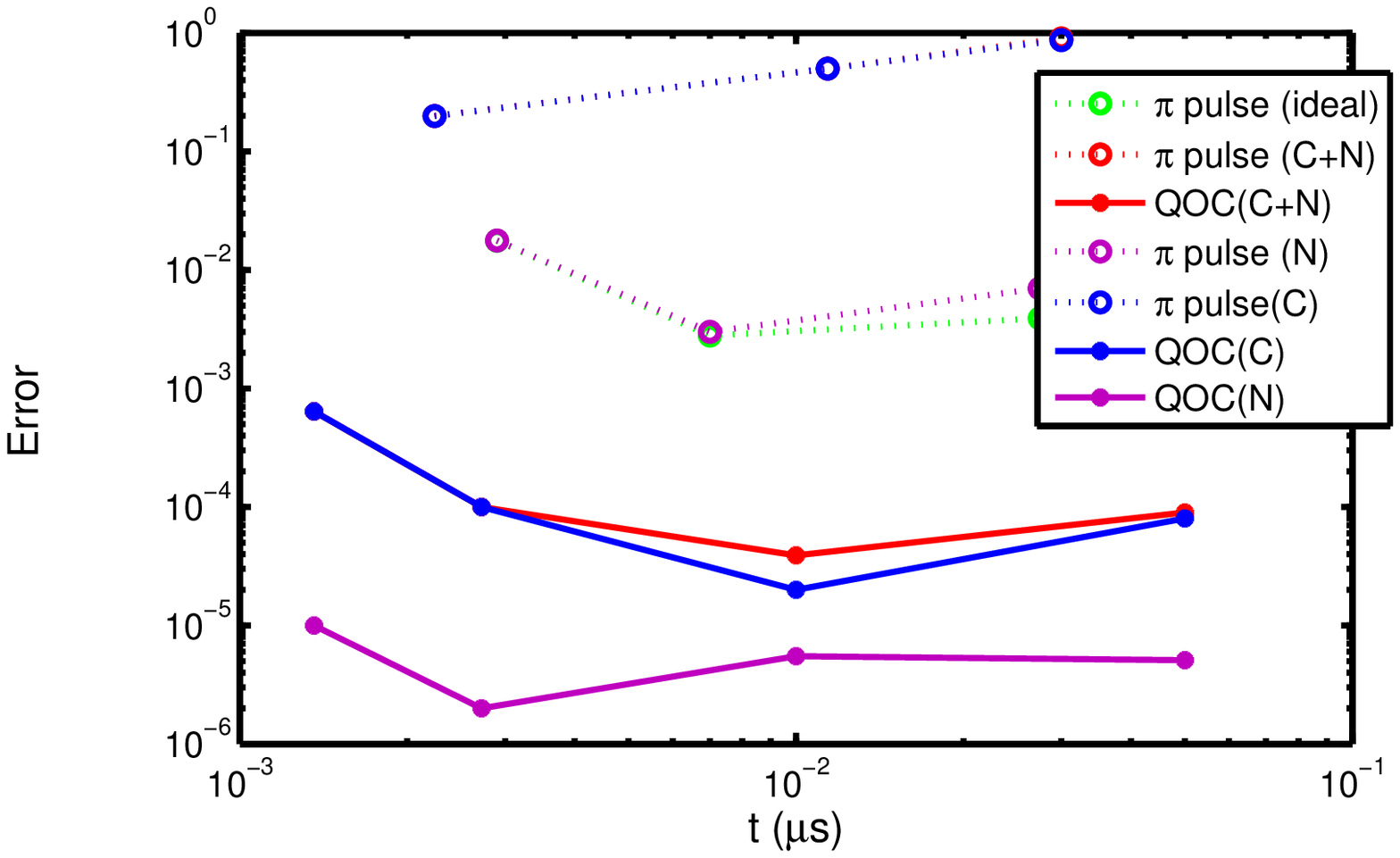}}
\subfigure[]{\includegraphics[scale=0.46]{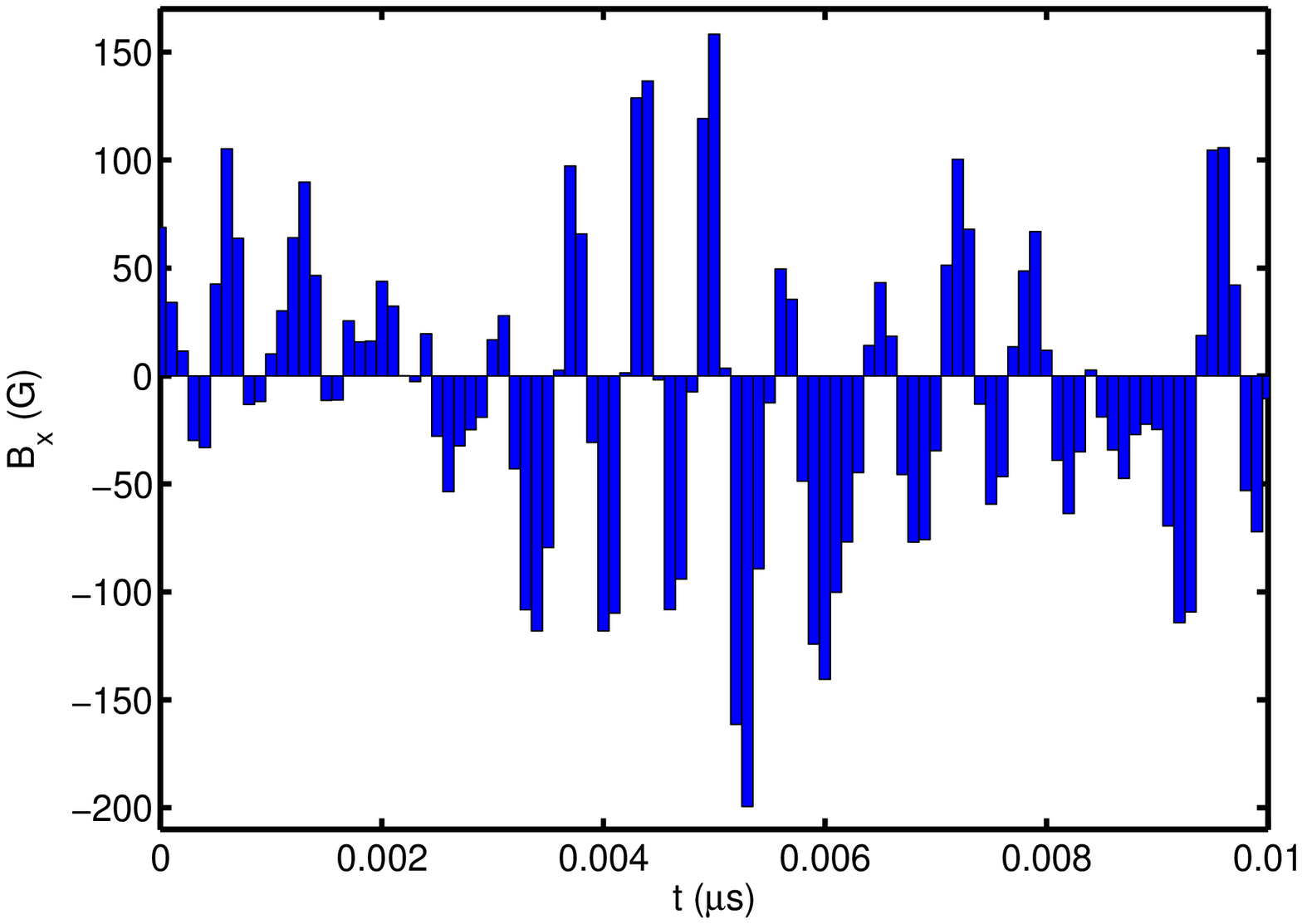}}
\caption{\label{fig:xgate_error} (Color online)
(a) $X$ gate errors versus operation times  at a static magnetic field 
$B_z = \mathrm{500 \; G}$ for different noise-qubit scenarios and
operation schemes.  The colors of purple, blue and red represent the cases
where the system qubit (electron spin) interacts with only the $\mathrm{^{15}N}$, only
the $\mathrm{^{13}C}$, and both the $\mathrm{^{15}N}$ and the
$\mathrm{^{13}C}$ noise qubits, respectively.  The open circles with dashed lines represent $X$
gates implemented by conventional $\pi$ pulse, while the solid squares
with solid lines denote implementation by optimal control pulse
sequence. The open green circles with
a dashed line stand for the $X$ gate error of an ideal NV electron
spin without any noise qubit nearby.
(b) A typical Optimal control pulse sequence of $B_x$ field for a gate
operation time of $\mathrm{0.01\;\mu s}$ at a static magnetic field of
$B_z = \mathrm{500 \; G}$.}
\end{figure}

For the implementation of an $X$ gate, traditionally 
a $\pi$ pulse with frequency $\omega$
in resonance to the energy splitting 
between the two computational states of the system
qubit (in our case $\omega=\mathrm{1.343\cdot 2\pi \; GHz}$
at $B_z = \mathrm{500 \; G}$) is used to
induce the qubit transition.
As fast quantum gates are favorable
for the purpose of 
quantum information processing, we set the operation times of
the single-qubit gates to be in the order of $0.01\mu$s.  
To be operated in such a short time, an $X$ gate implemented by a
$\pi$ pulse would require a certain pulse strength $B_{x0}$
of $B_x(t)=B_{x0}\cos{\omega t}$. This in
turn limits the fidelity of the gate as a relatively strong pulse
strength  $B_{x0}$ would
cause possible transitions to the $m_s=+1$ leakage state. One can see this from
Fig.~\ref{fig:xgate_error}(a) that the gate errors of $X$ gates
implemented by $\pi$ pulses for different operation
times in the absence of any noise qubit and decoherence
represented by the green circles (guided by the green dotted
line) are all larger than  $4\times 10^{-3}$.  
In the presence of the nearby noise qubits,
nonsecular parts of the hyperfine couplings $A_{\perp}^{eC}$ to the
system qubit will cause an additional error to the $X$ gate implemented by
a $\pi$ pulse. 
The purple, blue and red circles (dashed lines) in
Fig.~\ref{fig:xgate_error}(a) represent the errors of the $X$ gates
implemented by $\pi$ pulses at different operation times 
 for the cases
where the system qubit interacts with only the $\mathrm{^{15}N}$, only
the $\mathrm{^{13}C}$, and both the $\mathrm{^{15}N}$ and the
$\mathrm{^{13}C}$ noise qubits, respectively. 
One can see that the gate errors of the blue and red dashed
lines overlapping 
with each other are considerably larger than those of 
the purple and green dashed lines also
overlapping with each other in the short gating time regime but
deviating a little bit at large gating times. 
This is because the hyperfine interaction of the $\mathrm{^{15}N}$
nuclear spin with the system qubit 
is about one to two orders of
magnitude smaller than the control field strength or interaction
for the $X$ gates with operation times shown in Fig.~\ref{fig:xgate_error}(a),
and thus the presence of the $\mathrm{^{15}N}$
nuclear spin does not introduce substantial error in the $X$ gate. 
On the other hand, the hyperfine interaction of the $\mathrm{^{13}C}$
nuclear spin with the system qubit is comparable to the control field
interaction and thus considerably larger gate error is introduced by
$\mathrm{^{13}C}$ nuclear spin than that by the leakage state in the
ideal case or than that by the 
$\mathrm{^{15}N}$ nuclear spin. 

In contrast, the QOC based on the Krotov optimization method enables
us to achieve a 
high-fidelity $X$ gate with significant improvement in gate error. 
As shown in Fig.~\ref{fig:xgate_error}(a), the $X$ gate errors
obtained by the optimal control method in the
presence of both the $\mathrm{^{15}N}$ and the
$\mathrm{^{13}C}$ noise qubits represented by the red squares (solid
lines) are two orders of magnitude smaller than the gate error
obtained by 
directly applying $\pi$ pulses even in the ideal case (green dotted line). 
Figure \ref{fig:xgate_error}(b) shows a typical optimal control field
sequence for a $X$ gate with an operation time of $\mathrm{0.01\;\mu s}$.
We have also performed optimal control calculations for the single qubit
gates of the Hadamard gate ($H$ gate), the phase gate and the $\pi/8$
gate in the 
universal quantum gate set for FTQC. 
The errors of the single-qubit gates with operation times all set to
$\mathrm{0.01\;\mu s}$ are summarized in Table \ref{tab:errors}.
Because the gate times of the single-qubit gates 
are all set to a relatively short time of
$\mathrm{0.01\;\mu s}$,
we find that the calculated optimal control pulse sequences are robust 
(i.e., the gate errors do not increase appreciably) 
in the presence of a spin bath with a wide range of realistic
experimental parameters for the bath correlation function defined in 
(\ref{eq:bathCF}) \cite{raey,van-der-Sar2012,deLange01102010}.
The effect of a spin bath will be explicitly 
discussed for two-qubit
{\sc CNOT} gates that have longer gate times.

\begin{table}[htpb]\begin{center}
\caption{Summary of the QOC gate errors.
The bath correlation time
 for the {\sc CNOT} gates is $\tau_c=\mathrm{25\;\mu s}$, and the
average system-bath coupling strengths
(or field inhomogeneities) are $b=\mathrm{38.46\; \mu s^{-1}}$
for the gate time $T=\mathrm{0.125\;\mu s}$ and  
$b=\mathrm{80\; \mu s^{-1}}$ for the gate time $T=\mathrm{0.05\;\mu
  s}$, respectively.
\label{tab:errors}}
\begin{tabular}{llll}
\\
\hline \hline 
gate type & noise qubit(s) & gate time $T$ & gate error $K$  \\ \hline
\multirow{2}{*}{Z gate} & $\mathrm{^{15}N}$ & $\mathrm{0.01\;\mu s}$ & $8.0 \times 10^{-5}$ \\
 & $\mathrm{^{15}N},\mathrm{^{13}C} $ & $\mathrm{0.01\;\mu s}$ & $6.0 \times 10^{-4}$ \\ \hline
\multirow{2}{*}{X gate} & $\mathrm{^{15}N}$ & $\mathrm{0.01\;\mu s}$ & $5.5 \times 10^{-6}$ \\
 & $\mathrm{^{15}N},\mathrm{^{13}C} $ & $\mathrm{0.01\;\mu s}$ & $3.9 \times 10^{-5}$ \\ \hline
\multirow{2}{*}{H gate} & $\mathrm{^{15}N}$ & $\mathrm{0.01\;\mu s}$ & $2.4 \times 10^{-5}$ \\
 & $\mathrm{^{15}N},\mathrm{^{13}C} $ & $\mathrm{0.01\;\mu s}$ & $1.0 \times 10^{-4}$ \\ \hline 
\multirow{2}{*}{Phase gate} & $\mathrm{^{15}N}$ & $\mathrm{0.01\;\mu s}$ & $2.5 \times 10^{-5}$ \\
 & $\mathrm{^{15}N},\mathrm{^{13}C} $ & $\mathrm{0.01\;\mu s}$ & $4.3 \times 10^{-5}$ \\ \hline 
\multirow{2}{*}{$\pi /8$ gate} & $\mathrm{^{15}N}$ & $\mathrm{0.01\;\mu s}$ & $2.3 \times 10^{-5}$ \\
 & $\mathrm{^{15}N},\mathrm{^{13}C} $ & $\mathrm{0.01\;\mu s}$ & $1.4 \times 10^{-4}$ \\ \hline   
\multirow{2}{*}{CNOT gate} & $\mathrm{^{15}N}$ & $\mathrm{0.125\;\mu
  s}$ & $1.7 \times 10^{-4}$ \\ 
& $\mathrm{^{15}N}$, bath & $\mathrm{0.125\;\mu s}$ & $6.5 \times 10^{-4}$ \\ 
\hline
\multirow{2}{*}{CNOT gate} & $\mathrm{^{15}N}$ & $\mathrm{0.05\;\mu
  s}$ & $4.3 \times 10^{-4}$ \\ 
& $\mathrm{^{15}N}$, bath & $\mathrm{0.05\;\mu s}$ & $4.3 \times 10^{-4}$ \\ 
\hline \hline
\end{tabular}
 \end{center}\end{table}

\subsection{Optimal control for CNOT gates}

We next describe the implementation of the two-qubit CNOT gate in the
discrete set of universal gates. 
We choose the electron spin of the NV center as the control qubit and the $\mathrm{^{13}C}$
nuclear spin as the target qubit. The $\mathrm{^{15}N}$ nuclear spin
associated with the NV center acts as a noise qubit influencing the
electron qubit. 
Besides, the bath of other distant spin impurities causing the
decoherence to the 
electron qubit is taken into account. A schematic illustration of
the whole system we consider is shown in Fig.\ref{fig:model}(b).
The experiments carried out on a single electron and a single
$\mathrm{^{13}C}$ nuclear spin of a NV center for the implementation
of a  two-qubit conditional rotation 
({\sc CROT}) gate realized by a conditional radio-frequency $\pi$
pulse
have been reported
\cite{Jelezko2004,Shim2013}.
This {\sc CROT} gate combined with single-qubit $z$ rotations
can perform a {\sc CNOT}  gate up to a global phase factor.
Decomposing a {\sc CNOT} gate or
a general gate operation into several single-qubit and 
entangled two-qubit operations in series makes its operation
time normally longer 
and its overall gate error larger as the gate errors of the decomposed gates
will accumulate.
In contrast, the optimal control method has the great
advantage of enabling the implementation of a {\sc CNOT}  gate
or other general quantum gates in a single run of pulse
or in a single pulse sequence 
by simply setting the target
operation to be the {\sc CNOT} gate or the general quantum gate one
wishes to implement \cite{Tsai2009,Huang2014}. 
Simulating
this $\pi$-pulse approach of implementing the {\sc CROT} gate
using $B_x(t)={B_{x0}}\cos{\omega t}$ and
$B_y(t)={B_{yo}}\sin{\omega t}$ with the field
strengths $B_{x0}$ and $B_{y0}$, we evaluate the {\sc CROT} gate error $K$ and 
compare it with our result of {\sc CNOT} gate by the
optimal control method.

We discuss the implementation of a {\sc CNOT} or {\sc CROT} gate for
three different 
cases. The first case is the ideal case without any noise, the second
case includes the effect of the $\mathrm{^{15}N}$ noise qubit, and the
third case considers both the $\mathrm{^{15}N}$ noise qubit and a spin bath. 
Here applying a static magnetic field of $B_z = \mathrm{1000 \; G}$,
results in a relatively large energy splitting between
$m_s=1$ (leakage) state and $m=0,-1$ (computational basis) states
as compared to the energy splitting between the computational basis
states of 
the $m_s = 0$ and $m_s = -1$ states for the NV electron spin.
At first, the NV electron spin is regarded as an ideal two-level qubit
system for the gate implementation, and then 
we will treat the NV electron spin as a spin-1 three-level
system.  
This enables us to exam how well the optimal
control field sequence obtained for the two-level case performs in the more
realistic three-level case (i.e., to see the leakage effect).
We then employ the QOC
theory to find new control sequence for the three-level NV center
electron spin taking the leakage state of $m_s=1$ into account 
in order to reduce the gate error or infidelity.

\begin{figure}
\centering  
\includegraphics[scale=0.48]{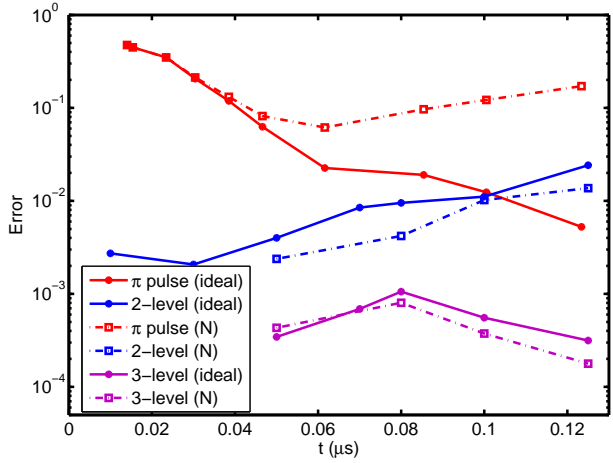}
\caption{\label{fig:error} (Color online)
{\sc CNOT} or {\sc CROT} gate errors versus operation times  at a
static magnetic field $B_z = \mathrm{1000 \; G}$ for different
operation schemes: Conditional $\pi$ pulse scheme (in red)
and optimal control scheme taking the NV
center electron spin as a two-level system (in blue) and three-level
system (in purple).
The solid lines and dot-dashed lines represent the gate errors obtained for
the ideal case and the case including noise qubit $\mathrm{^{15}N}$,
respectively.}
\end{figure}

{\em Ideal case:} Next we show that the optimal control theory can
achieve a {\sc CNOT} gate with a better fidelity than simply applying
a conditional $\pi$ pulse ({\sc CROT}).  
The {\sc CNOT} or {\sc CROT} gate errors as a function of the gate
operation time for 
different cases are illustrated in Fig.~\ref{fig:error}. 
The solid lines and dot-dashed lines represent the gate errors obtained for
the ideal case and the case including noise qubit $\mathrm{^{15}N}$,
respectively. The red lines 
represent the {\sc CROT} gate errors
obtained by applying a conditional $\pi$ pulse taking $m_s=0,-1$ states
of the NV center as the system qubit states.
The blue lines and purple lines represent the {\sc CNOT} gate
errors obtained by using the optimal control theory taking the NV
center electron spin as a two-level system and three-level system,
respectively. Note that different from the gate error of the CROT
implemented by a conditional $\pi$ pulse, the QOC gate errors are
calculated by substituting 
the pulse sequence obtained in respective methods into the
Hamiltonian of 
the three-level NV center system 
including the $\mathrm{^{15}N}$ noise qubit
(but not including the bath spins) for the
propagator $U$ in Eq.~(\ref{eq:error_leakage}). 
For the idea case (solid lines), 
when the gate operation time becomes shorter, the gate error using a
$\pi$ pulse becomes larger. The reason is that the 
requirement for a shorter duration of a $\pi$ pulse means a stronger
field strength, which in turn induces a larger transition probability to
the leakage state. In contrast, the QOC theory can do a
much better job when the operation time is short as illustrated in
Fig.\ref{fig:error}. However, as the operation time increases, the
error for the optimal control case of treating the electron spin as an
effective two-level system becomes larger, even larger than that of
using a $\pi$ pulse (see Fig.\ref{fig:error}). 
Therefore, employing QOC for the three-level
electron spin system is necessary.
It is quite obvious from 
Fig.\ref{fig:error} that employing
the QOC theory by treating the electron spin as a
three-level system (purple solid lines) gives gate errors one order to
two orders of magnitude
lower than
those of the two-level case (blue solid line). 


\begin{figure}
\centering
\subfigure[]{\includegraphics[scale=0.58]{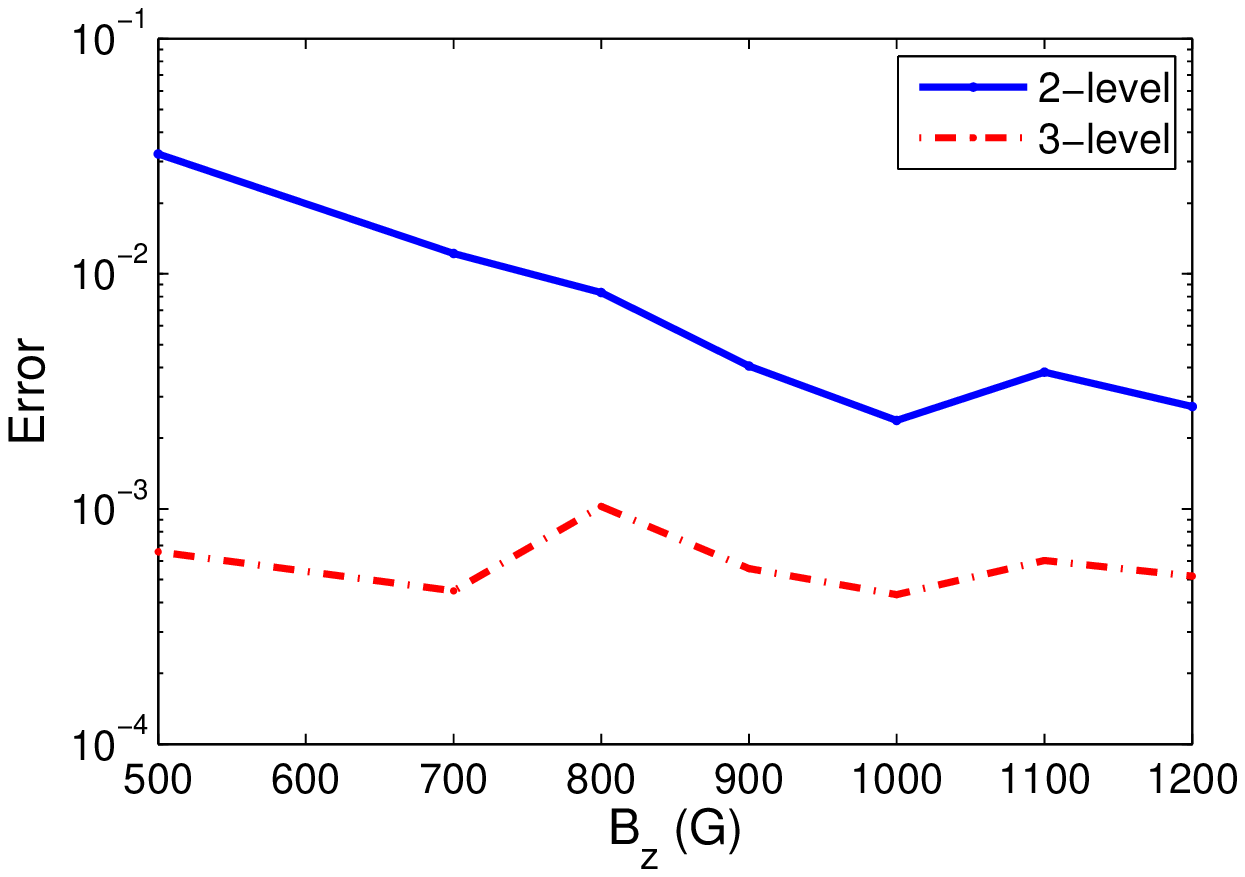}}
\subfigure[]{\includegraphics[scale=0.5]{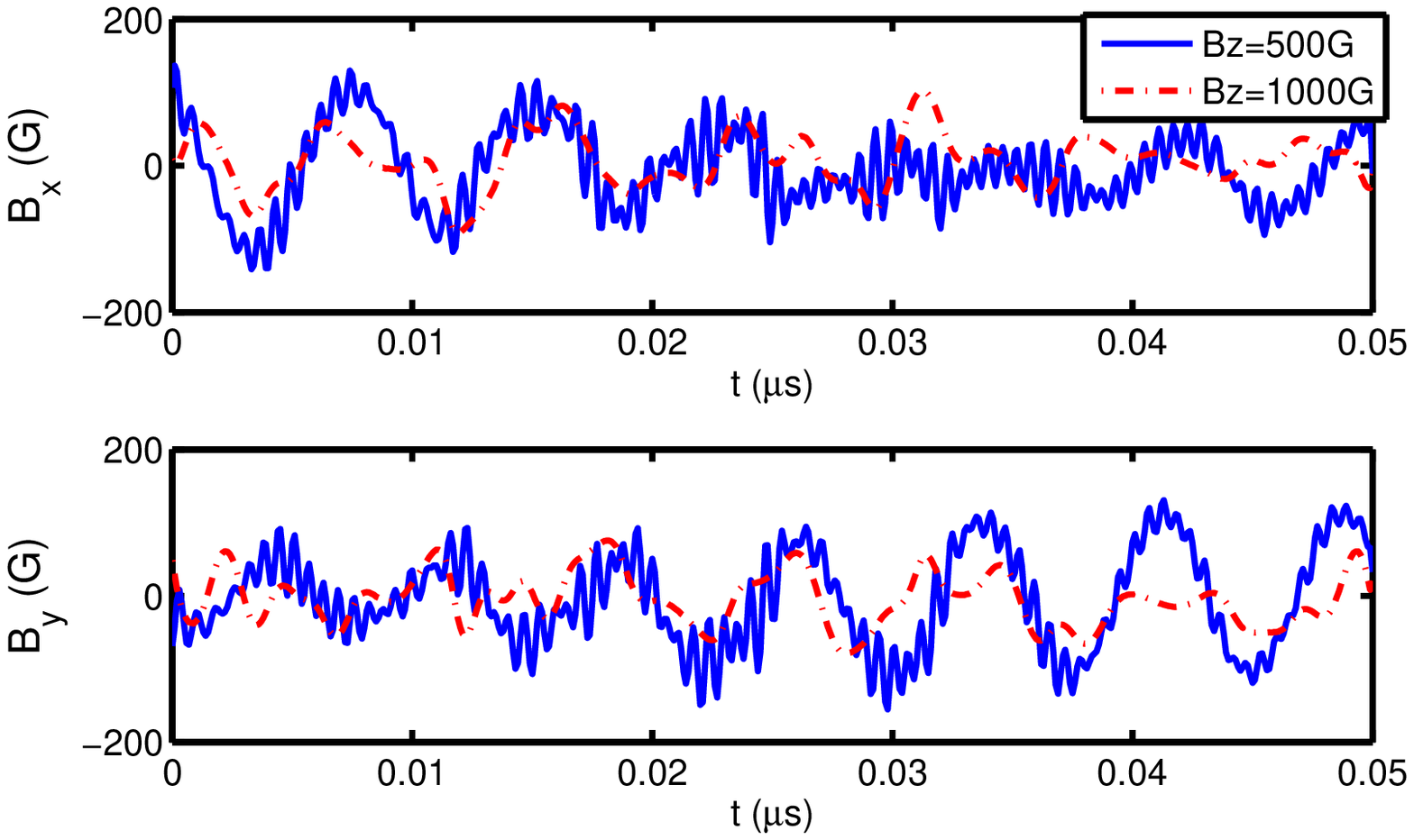}}
\caption{\label{fig:error_Bz} (Color online) 
(a) Optimal control {\sc CNOT} gate errors for different values of the
static magnetic field $B_z$ taking the NV
center electron spin as a two-level system (blue solid line) and three-level
system (red dot-dashed line). 
The operation time is $\mathrm{0.05\;\mu s}$. 
(b)Optimal control pulse sequences of $B_x(t)$ and $B_y(t)$ for $B_z =
\mathrm{500 \; G}$ (blue solid lines) and $B_z = \mathrm{1000 \; G}$
(red dot-dashed lines)}.
\end{figure}

{\em Effect of a noise qubit:} 
In the case of including the $\mathrm{^{15}N}$ noise qubit, driving the
system by a $\pi$ 
pulse does not work well anymore (see the red dot-dashed
line). 
However, taking the noise qubit into the QOC optimization consideration lowers the error or
infidelity a little bit as compared to the case without doing so. One
can see this from Fig.\ref{fig:error} that
the blue and purple dot-dashed
lines are slightly lower than their corresponding blue and purple solid
lines, respectively. 
The reason is that the $^{15}{\rm N}$ noise qubit here also serves as an
ancilla qubit that allows some probability to get out of the 
computational space temporarily but return to the computational space
with higher fidelity at the end of the operation for the optimal
control {\sc CNOT} gate. 　 
Again, treating the electron spin as a three-level system
is a more effective
strategy to perform {\sc CNOT} gate and it gives an error $K\approx 1.7\times
10^{-4}$ in the presence of the $^{15}{\rm N}$ noise qubit for a gate time of 
$\mathrm{0.125\;\mu  s}$.
Figure \ref{fig:error_Bz}(a) investigates the optimal control {\sc CNOT}
gate error as a 
function of the strength of the external static magnetic field $B_z$ for
a gate operation time of $\mathrm{0.05\;\mu s}$. The blue solid line
and red dot-dashed line represent the cases of 
treating the electron spin as a two-level system and a three-level
system, respectively. 
When the static magnetic field $B_z$ decreases, the gate error 
represented by the blue solid line becomes large. 
This is because at a lower magnetic
field $B_z$ the energy separation between the qubit state $m_s=-1$ and
the leakage state $m_s=1$ become smaller, and thus the approximation of
treating the NV electron spin as a two-level system of $m_s=0,-1$ is not
very good. 
However, the optimal
control of the three-level case still works with error $K\approx
6\times 10^{-4}$  even when the magnetic field is as low as
$B_z=\mathrm{500\;G}$. 
The optimal control field sequences by treating the NV electron spin
as a three-level system for $B_z=1000$G (red dot-dashed lines) and
$B_z=500$G (blue solid lines)
with the operation time of $\mathrm{0.05\;\mu s}$ 
are shown in Fig.~\ref{fig:error_Bz}(b). 
At $B_z=\mathrm{500\; G}$ ($B_z=\mathrm{1000\; G}$), the energy
separation between the electron qubit states of $m_s=-1$ and  $m_s=0$
is much larger than (comparable to) 
the energy splitting between the
states $\ket{1,\downarrow}$ and $\ket{1,\uparrow}$ (due to the hyperfine
interaction), where
$\ket{\downarrow}$ and $\ket{\uparrow}$ represent the
$\mathrm{^{13}C}$ nuclear spin states, and $\ket{1}$ and $\ket{0}$
represent the electron qubit states.
As a result, there are apparently two different oscillating 
components for the case of $B_z=500$G (blue solid lines) as compared to
the case of $B_z=\mathrm{1000\; G}$ (red dot-dashed lines) in
Fig.~\ref{fig:error_Bz}(b). The major frequency of the fast oscillating
component comes from the high frequency (energy
separation) of the electron qubit, while that of the slow one matches the
energy splitting between 
states $\ket{1,\downarrow}$ and $\ket{1,\uparrow}$. 

\begin{figure}
\centering
\subfigure[]{\includegraphics[scale=0.5]{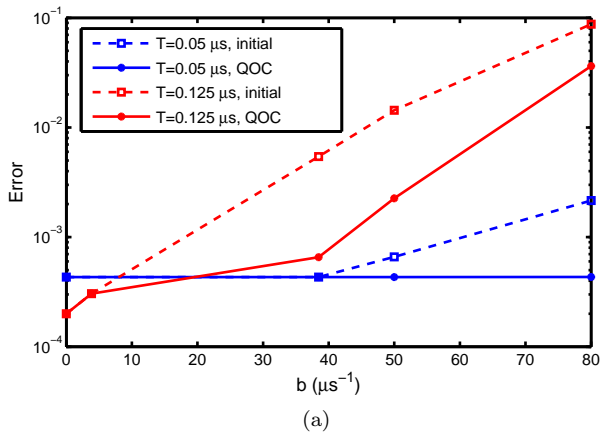}}
\caption{\label{fig:error_b} (Color online)
{\sc CNOT} gate error verse the average system-bath coupling strength (field
inhomogeneity) $b$ for gating time of $=\mathrm{0.05\;\mu s}$ (blue
solid line) and $\mathrm{0.125\;\mu s}$ (red solid line). 
The corresponding dashed lines represent the gate errors calculated by taking 
the OCT pulse sequences obtained for the case without the spin bath
(i.e., setting $b=0$)
as our initial guess for optimal control in the open system.
} 
\end{figure} 

{\em Effect of a spin bath:}
Next, we take into account not only the noise qubit but also the bath of the
distant spin impurities, which is modeled as an effective classical
random field acting on the NV electron spin with correlation function
given by Eq.~(\ref{eq:bathCF}). 
The solid lines in Fig.~\ref{fig:error_b} show the dependence of the
{\sc CNOT} gate error　 
on the bath parameter $b$ for the correlation time
$\tau_c=\mathrm{25\; \mu s}$. The parameter $b$ defined in the bath
correlation function, Eq.~(\ref{eq:bathCF}), is related to the average 
system-bath coupling strength or field inhomogeneity in the random
field model. The dashed lines in
Fig.~\ref{fig:error_b} represent the gate error obtained by applying the optimal
control sequences obtained for the case without the spin bath to the
master equation or the effective evolution equation
(\ref{eq:motion_open_U}) in the presence of the bath. 
This allows us to study how the gate errors due to the spin bath are
improved in the open system by QOC. 
One can see from Fig.~\ref{fig:error_b} that for $\tau_c=\mathrm{25\; \mu s}$
unless the parameter $b$ is more than one order of magnitude
larger than the typical value of $b=\mathrm{3.846\; \mu s^{-1}}$ extracted
from the experiments \cite{raey,van-der-Sar2012,deLange01102010}, the
influence of the bath on the gate with a short
operation time of $\mathrm{0.05\;\mu s}$ (blue dashed line) is not significant. 
For the
longer gate operation time of $\mathrm{0.125\;\mu s}$ red dashed line), the 
influence of a bath 
becomes 
appreciable when $b \gtrsim \mathrm{5 \; \mu s^{-1}}$, and  
as the coupling strength $b$ become stronger, the error
increases. 
However, the optimal control {\sc CNOT} gate with operation time
$\mathrm{0.05\;\mu s}$ 
(blue solid line)
can sustain much larger values of the field inhomogeneity
up to $b=80\; \mu s^{-1}$ and the gate error can be maintained almost the
same as if the spin bath were not present.   

\begin{figure}
\centering
\subfigure[]{\includegraphics[scale=0.46]{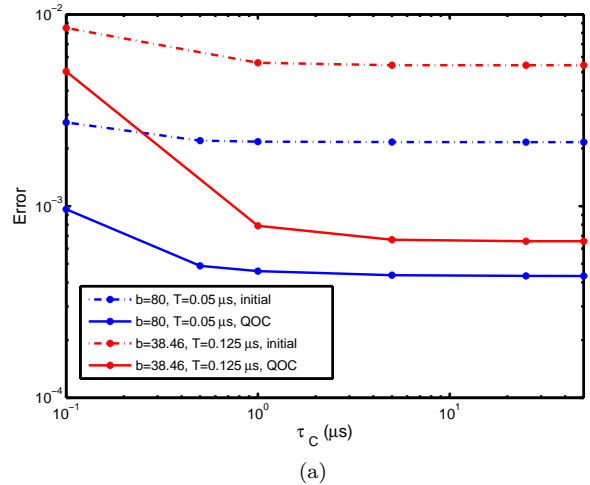}}
\caption{\label{fig:error_tau} (Color online)
{\sc CNOT} gate error verse the correlation time
  $\tau_c$ for the case of $b=\mathrm{38.46\; \mu s^{-1}}$
and the gating time $T=\mathrm{0.125\;\mu s}$ and the case of 
$b=\mathrm{80\; \mu s^{-1}}$ and the gating time $T=\mathrm{0.125\;\mu
  s}$) 
The dashed line represents gate error calculated by taking 
the OCT pulse sequences obtained for the case without the spin bath
(i.e., setting $b=0$)
as our initial guess for optimal control in the open system.}
\end{figure} 

The relation between the gate error and the bath correlation time
$\tau_c$ for the average coupling strengths of $b=\mathrm{38.46\; \mu s^{-1}}$
and the operation time of $\mathrm{0.125\;\mu s}$ 
($b=\mathrm{80\; \mu s^{-1}}$ and operation time $\mathrm{0.125\;\mu
  s}$) is shown in red solid line (blue solid line) in
Fig.~\ref{fig:error_tau}. 
To investigate how much QOC theory
improves the gate fidelity in the NV center system with a spin bath,
we take
the optimal pulse sequences in the absent of the spin bath as our initial
guess for the control fields, and the gate errors before the optimal
control iterations are shown in dot-dashed lines in Fig.\ref{fig:error_tau}.  
The gate  
errors (solid lines) after the iterations are terminated are larger
for small values of $\tau_c$, and decrease as $\tau_c$
increases. The gate errors (solid lines)
stay nearly constant as
$\tau_c \gtrsim  \mathrm{5\; \mu s}$. 
When the bath correlation time  $\tau_c$ is small,
the memory effect of the bath is weak and the bath is close to a Markovian
bath. In this case, it is hard to revise the bath contribution since the decay
rate approaches to a steady value in a very short time. 
In contrast,  when $\tau_c$ is large,
due to the relatively long  
memory effect of the bath, 
the optimal
control fields are able to counteract the influence from the bath.

\section{CONCLUSION}\label{sec:conclusion}

We have found the control sequences of fast and high-fidelity
single-qubit and two-qubit quantum gates for electron and nuclear
spins of a NV center in 
diamond using the QOC theory. 
A {\sc CNOT} gate or other general quantum gates operation can be
implemented in a single run of pulse sequence using the optimal
control approach rather than being
decomposed into some entangled two-qubit and several single-qubit 
operations in series by composite pulse sequences.
The (non-Markovian) external environment effects
on system qubits are partitioned into two kinds, one from a few nearby
noise qubits and the other from a bath of distant spins.
Table \ref{tab:errors} summarizes the gate time and gate
errors calculated with realistic experimental parameters for
the cases considering the effect 
of the noise qubits, the leakage state and the effect of a spin bath.
These gate errors are
below the recent model of error threshold $10^{-3}$
\cite{Aliferis2009} ($10^{-2}$ if surface code error
correction is used \cite{Wang11,Fowler12L,Fowler12}) 
required for FTQC. 
One can estimate the logical error rate that our gate error $K$
corresponds to for a given type of error correction of FTQC. 
Suppose that one implements a logical qubit 
of the surface code error correction on a
two-dimensional array of physical qubits with
array size (distance) $d$  [corresponding to the number
of physical qubits being $n_q=(2d-1)^2$] as in Ref.~\cite{Fowler12}.
The number of physical qubits or the array size $d$ needed to
define a logical qubit to meet a required logical error rate 
is dependent strongly on the error rate in the physical qubits.
By taking the error probability per step to be the worse gate error $K$ of
$6.0\times 10^{-4}$ in Table \ref{tab:errors}, which is smaller than
the error threshold rate of $p_{\rm th}=0.57\%$ of the surface code,
the estimated logical 
error rates using Fig.~4 or more precisely Eq.~(11) of Ref.~\cite{Fowler12} are
about $3.5\times 10^{-5}$ for array size $d=5$ and about $3.7\times 10^{-6}$ 
for array size $d=7$. When the per-step error rate is smaller than the error
threshold rate $p_{\rm th}$, the logical error rate falls exponentially with
the array size $d$. Thus one obtains the
logical error rate for array size $d=25$ to be about $5.8\times
10^{-15}$, sufficient to perform Shor's algorithm for factoring a
$2000$-bit number into its primes  
with a reasonable chance of success \cite{Fowler12}.

There seems to be challenges for the experimental implementations and
applications of the QOC theory, such as imprecise knowledge of the
quantum system's parameters and how to generate the complex optimal control
pulses in reality. 
Fortunately, commercial
devices for generating arbitrary
wave forms or complex signals in a time scale of sub-nanoseconds to
nanoseconds are available now and may solve the challenge of
generating complex pulse sequences.
Besides, a hybrid open-loop--closed-loop optimal control method called
adaptation by hybrid optimal control (Ad-HOC) method \cite{Egger14} designed to
overcome not only the problem of inaccurate knowledge of the system
parameters but also shortcomings of the assumed physical model and
errors on the control fields has been recently proposed.
The closed-loop pulse calibration of Ad-HOC,
similar to adaptive model-free feedback control (also referred
to as closed-loop laboratory control or learning control)
\cite{Rabitz1992,Rabitz2010}, uses the physical system itself as a feedback to
calibrate control pulses and optimize their performance.
Two similar closed-loop methods for optimizing quantum control in experimental systems
have also been put forward recently: the method of optimized randomized
benchmarking for immediate tune-up (ORBIT) \cite{Kelly14} and the 
method of adaptive control via randomized optimization nearly
yielding maximization (ACRONYM) \cite{Ferrie14}.
In principle, the system response and control pulses can be calibrated
and improved using closed-loop optimization where measurement data are
efficiently obtained with Nelder-Mead algorithm \cite{Egger14,Kelly14,Nelder1967} or stochastic
optimization algorithm \cite{Ferrie14,Spall1992,Spall2005} and 
subsequently fed back to the system optimizer to improve the pulses without
precise knowledge of the system. 
These developments make the QOC theory practical and useful to construct the
initial pulse sequences for experimentally closed-loop optimization
\cite{Egger14}.

\begin{acknowledgments}
We acknowledge support from the the
Ministry of Science and Technology of Taiwan under
Grant No.~103-2112-M-002-003-MY3,  from
the National Taiwan University under Grants No.~NTU-ERP-104R891402,
and from the thematic group program of the 
National Center for Theoretical Sciences, Taiwan. 

\end{acknowledgments}

\appendix
\section{ Explicit form of $dK/dU$} \label{sec:Append}

Employing the algorithm of the Krotov optimization method, we evolve
backward in time the auxiliary function 
$\mathcal{B}(t)$
with the boundary condition $\mathcal{B}(T)=-\frac{dK}{dPU}
$. Therefore, we need to find the explicit form of
$\frac{dK}{dPU}$. The procedure and calculation presented below follows those
 in Ref.~\onlinecite{0953-4075-40-9-S06}.

In our case, the  error function $K$ is defined through
Eqs.~(\ref{eq:error_leakage}) and (\ref{eq:Q}).
The matrix elements of $\frac{dK}{dPU}$ can be written as 
\begin{equation}
 \left( \dfrac{dK}{dPU} \right)_{ab}=\dfrac{dK}{dPU_{ab}}, 
\end{equation}  
where the matrix indices $a$ and $b$ range form $1$ to $N$ and
$PU_{ab}$ is the matrix element (a complex scalar variable) of $PU$.
Let $x=PU_{ab}$ , and $Z(x)=\sqrt{Q^\dagger Q}$ be a matrix function
of the variable $x$, and $y(Z(x))=\mathbf{Tr}\sqrt{Q^\dagger
  Q}=\mathbf{Tr}Z(x)$ be a scalar function. 
Thus, the error function (\ref{eq:error_leakage}) in terms of the new
variables is $K=\frac{1}{2}+\frac{1}{2N}\mathbf{Tr}[(PU)^\dagger
PU]-\frac{1}{N}y$.
Note that $\sqrt{Q^\dagger Q}$ is not an analytic function of $PU_{ab}$
but can be expressed as an analytic function of $PU_{ab}$ and
$PU_{ab}^*$. Consequently, when one differentiates $\sqrt{Q^\dagger
  Q}$ with respect to $PU_{ab}$, $PU_{ab}^*$ and subsequently
$Q^\dagger$ are treated as constants.  Thus 
\begin{align}
\dfrac{dK}{dx}&= \frac{1}{2N} PU_{ba}^*- \dfrac{1}{N}\dfrac{dy}{dx}\nonumber \\
&= \frac{1}{2N} PU_{ba}^*- \dfrac{1}{N}\sum_{k,k'} \dfrac{dy}{dZ_{k,k'}}\dfrac{dZ_{k,k'}}{dx} \nonumber \\
&= \frac{1}{2N} PU_{ba}^*-\dfrac{1}{N}\sum_{k,k'} \dfrac{dy}{dZ_{k,k'}}\dfrac{dZ_{k',k}^T}{dx} \nonumber \\
&= \frac{1}{2N} PU_{ba}^*-\dfrac{1}{N} \mathbf{Tr}\left(  \dfrac{dy}{dZ}\dfrac{dZ^T}{dx} \right) \nonumber\\
&= \frac{1}{2N} PU_{ba}^*-\dfrac{1}{N} \mathbf{Tr}\left( \dfrac{dZ^T}{dx} \right). \label{eq:dKdU}
\end{align} 
Here the second line in Eq.~(\ref{eq:dKdU}) follows from the use of
the chain rule, and the last line follows from the property of
$\dfrac{dy}{dZ} =\dfrac{d\mathbf{Tr}Z(x)}{dZ}=I$.  
Then from the definition of $Z$ and $x$, we find that
\begin{equation}
\dfrac{dZ}{dx}=\dfrac{d\sqrt{Q^\dagger Q}}{dPU_{ab}}=\frac{1}{2}(Q^\dagger Q)^{-1/2} Q^\dagger \dfrac{dQ}{dPU_{ab}}. \label{eq:dzdx}
\end{equation} 
Since $\mathbf{Tr}\dfrac{dZ^T}{dx}= \mathbf{Tr}\dfrac{dZ}{dx}$, 
inserting the trace of Eq.~(\ref{eq:dzdx}) into Eq.~(\ref{eq:dKdU}), we finally obtain the explicit form of 
  \begin{equation}
\left( \dfrac{dK}{dPU} \right)_{ab}=\frac{1}{2N}\left\{ PU_{ba}^*- \mathbf{Tr} \left[ (Q^\dagger Q)^{-1/2} Q^\dagger \dfrac{dQ}{dPU_{ab}}\right]\right\},
\label{eq:dKdPU}
\end{equation}
where 
\begin{equation}
\dfrac{dQ}{dPU_{ab}}=G^*_{i_a j_b}\ket{a \;{\rm mod}\; n_B}\bra{b\; {\rm mod}\; n_B} \label{eq:dQdU}
 \end{equation}
can be simply obtained by the definition of $Q$ in Eq.~(\ref{eq:Q}).
Here $i_a$ and $j_b$ 
denote the smallest integers greater than
or equal to 
$a/n_B$ and $b/n_B$, respectively, and $G_{i_a j_b}$ are the
matrix elements of the target gate. In Eq.~(\ref{eq:dQdU}), the states
$\ket{a\; {\rm mod}\; n_B}\bra{b\; {\rm mod}\; n_B}$ 
are the elements of the orthonormal basis matrix of the noise qubit(s).

\bibliographystyle{apsrev4-1}
\end{document}